\documentclass[12pt]{elsarticle}

\usepackage[ left=2.5cm, right=2.5cm]{geometry}

\usepackage{amssymb}
\usepackage{booktabs}
\usepackage{graphicx}
\usepackage{multirow}
\usepackage{tabularx} 
\usepackage{ragged2e}
\usepackage{tablefootnote}
\usepackage{amsmath}
\usepackage{float}
\usepackage{booktabs}
\usepackage{array} 
\usepackage[colorlinks=true, allcolors=blue]{hyperref} 
\usepackage{xcolor}
\usepackage{colortbl}  

\journal{Nuclear Physics B}

\begin{document}
    \begin{frontmatter}


        \title{OralGPT: A Two-Stage Vision-Language Model for Oral Mucosal Disease Diagnosis and Description}


        \author[1]{Jia Zhang\fnref{contrib}} 
        
        \ead{dentist.dj@xjtu.edu.cn}
        \author[2]{Bodong Du\fnref{contrib}}
        \ead{bduag@connect.ust.hk}
        \author[1]{Yitong Miao\fnref{contrib}}
        \ead{2204112921@stu.xjtu.edu.cn}
        \author[1]{Dongwei Sun \corref{cor1}}
        \ead{sundongwei@outlook.com}
        \author[1]{Xiangyong Cao \corref{cor1}}
        \ead{caoxiangyong@mail.xjtu.edu.cn}

        \affiliation[1]{organization={Xi’an Jiaotong University},
         addressline={ Xingqing Rd}, city={Xi'An}, postcode={710049 }, state={Shaanxi}, country={China}
        }
        \affiliation[2]{organization={The Hong Kong University of Science and Technology}
        , addressline={Clear Water Bay}, city={Kowloon}, postcode={HKG}, state={Hong Kong}, country={China}
        }
        \fntext[contrib]{These authors contributed equally.} 
        \cortext[cor1]{Corresponding author.} 



        \begin{abstract}
            Oral mucosal diseases such as leukoplakia, oral lichen planus, and recurrent
            aphthous ulcers exhibit diverse and overlapping visual features,
            making diagnosis challenging for non-specialists. While vision-language
            models (VLMs) have shown promise in medical image interpretation,
            their application in oral healthcare remains underexplored due to
            the lack of large-scale, well-annotated datasets. In this work, we present
            \textbf{OralGPT}, the first domain-specific two-stage vision-language
            framework designed for oral mucosal disease diagnosis and captioning.
            In Stage 1, OralGPT learns visual representations and disease-related
            concepts from classification labels. In Stage 2, it enhances its language
            generation ability using long-form expert-authored captions. To
            overcome the annotation bottleneck, we propose a novel similarity-guided
            data augmentation strategy that propagates descriptive knowledge from
            expert-labeled images to weakly labeled ones. We also construct the
            first benchmark dataset for oral mucosal diseases, integrating multi-source
            image data with both structured and unstructured textual annotations.
            Experimental results on four common oral conditions demonstrate that
            OralGPT achieves competitive diagnostic performance while generating
            fluent, clinically meaningful image descriptions. This study
            provides a foundation for language-assisted diagnostic tools in oral
            healthcare.
        \end{abstract}

        \begin{graphicalabstract}
            \includegraphics[width=\linewidth]{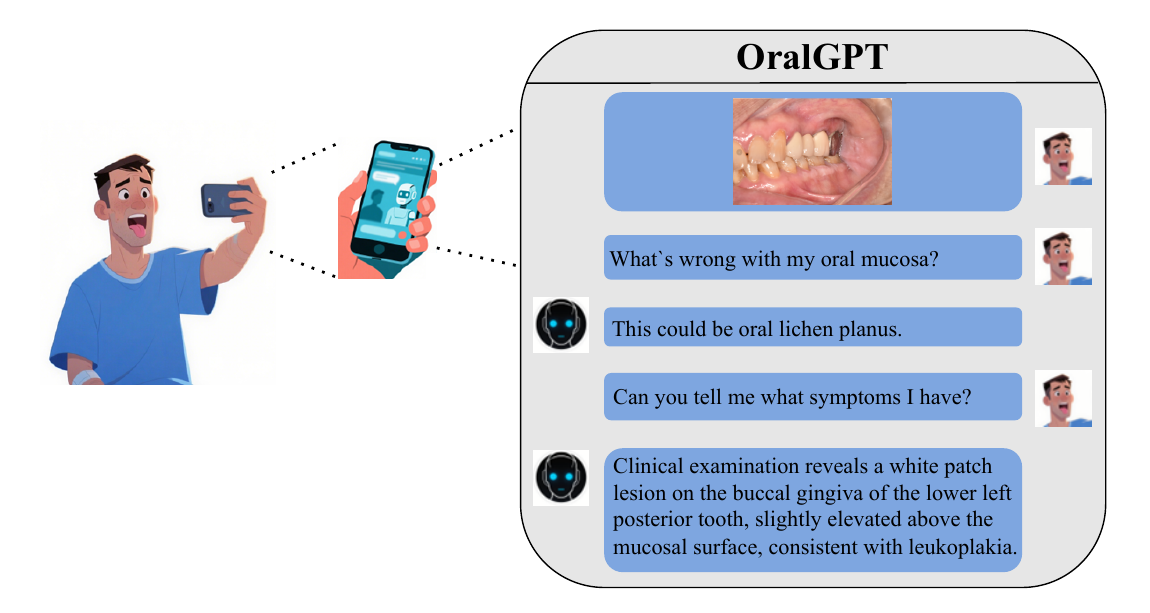}
        \end{graphicalabstract}

        \begin{highlights}
            \item We propose OralGPT, a two-stage vision-language framework for
            oral mucosal disease diagnosis and captioning under low-supervision
            conditions. 
            \item A similarity-guided pseudo-captioning mechanism enables
            reliable natural language supervision from weakly labeled clinical images,
            improving both classification and caption quality.
            \item We present the first benchmark for oral mucosal diseases with multi-source images and expert annotations.
        \end{highlights}

    \end{frontmatter}



\section{Introduction}

Oral mucosal diseases, including \textbf{Oral leukoplakia}, \textbf{Oral lichen planus}, \textbf{Recurrent oral aphthous ulcer} and \textbf{Discoid lupus erythematosus} \cite{intro1,intro2}, present with diverse and often
overlapping clinical manifestations, making accurate diagnosis challenging for non-specialists \cite{neville2023oral,warnakulasuriya2021oral}. These conditions affect millions globally, with oral leukoplakia demonstrating malignant transformation rates of 0.13-34\% \cite{lind1985malignant}, while oral lichen planus affects 1-2\% of adults with potential for malignant transformation \cite{bombeccari2011oral}. Early identification is critical to preventing malignant transformation \cite{intro3}.

Recent advances in \textbf{vision-language learning} have shown great
potential for enhancing medical image interpretation \cite{zhang2022contrastive,liu2023visual}. \textbf{Vision-language models (VLMs)} \cite{chen2024internvl,radford2021learning} offer a promising
pathway toward automated, interpretable diagnostic tools providing both classification results and clinical explanations. These models have demonstrated success in dermatology \cite{han2020augmented}, radiology \cite{pelka2018radiology}, and ophthalmology \cite{gulshan2016development,de2018clinically}. However, their
application in the \textbf{oral healthcare domain remains largely unexplored} \cite{heo2021artificial,hung2019application},
primarily due to the absence of large-scale datasets pairing oral lesion
images with high-quality, clinical descriptions.

The oral cavity presents distinct challenges for computer vision applications. Intraoral images exhibit substantial variability in lighting conditions, camera angles, and image quality \cite{zhu2023artificial}. Additionally, oral lesions manifest with subtle color variations, irregular borders, and heterogeneous surface textures that demand specialized preprocessing techniques \cite{khanagar2021developments,uthoff2018point}.

Automated oral disease diagnosis faces challenges related to dataset heterogeneity and standardization. Unlike radiology with established imaging protocols, oral medicine lacks comprehensive structured reporting frameworks \cite{rasteau2022artificial}. This results in inconsistent terminology usage and diverse annotation styles across clinical centers \cite{lopez2022machine}. The scarcity of large-scale, well-annotated oral disease datasets has limited machine learning model development \cite{aubreville2017automatic}.

\begin{figure}[htbp]
        \centering
        \includegraphics[width=1\linewidth]{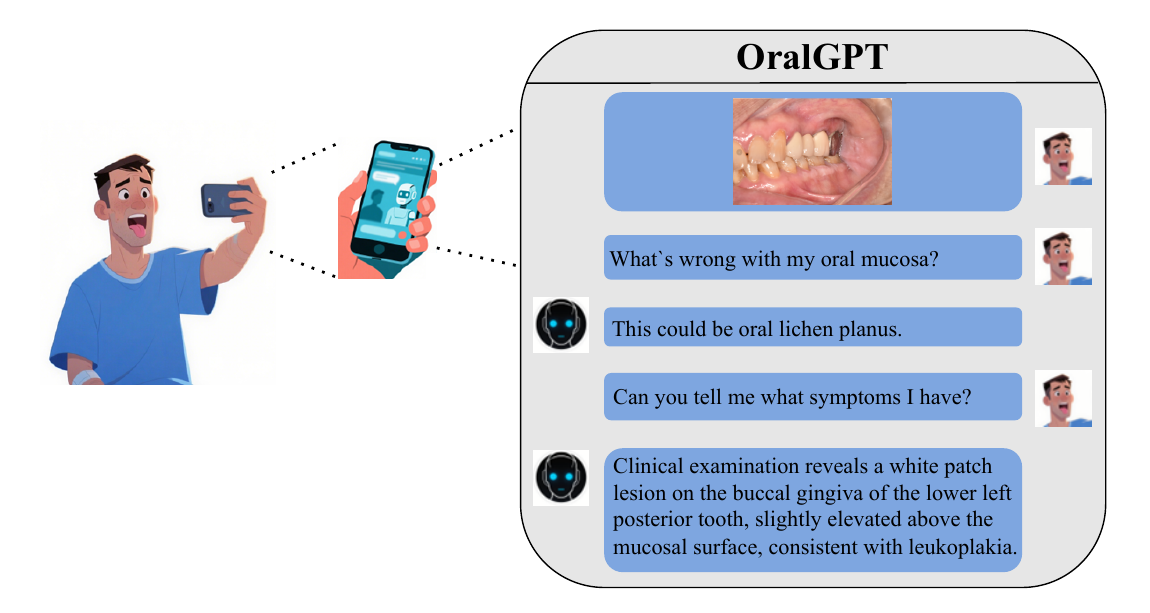}
        \caption{\textbf{Illustration of OralGPT.} OralGPT is an interactive oral mucosal disease diagnostic system based on multimodal large language models. We designed a two-stage vision-language framework that aligns a pre-trained vision transformer with Qwen2.5-VL-7B-Instruct. OralGPT was trained on a large-scale dataset containing four common oral mucosal diseases, integrating multi-source image data with expert annotations. Users can upload oral photos for diagnosis, and OralGPT can autonomously determine lesion characteristics and categories, provide clinical descriptions, and enable interactive diagnosis.}
        \label{tab:OralGPT}
    \end{figure}

    To bridge this gap, we propose \textbf{OralGPT} (Figure~\ref{tab:OralGPT}),
    a novel two-stage domain-specific vision-language learning framework
    tailored for oral mucosal disease diagnosis and description. Our study makes
    the following key contributions:

    \begin{itemize}
        \item \textbf{We establish the first benchmark dataset} for oral mucosal
            diseases that integrates multi-source image data with expert-annotated
            and weakly labeled descriptions, laying a foundational resource for future
            research in vision-language modeling within oral medicine.

        \item \textbf{We develop OralGPT}, the first domain-specific two-stage VLM
            framework for oral healthcare applications. In the first stage, OralGPT
            learns disease-related visual concepts and discriminative features from
            classification labels to build a solid foundation in visual understanding.
            In the second stage, it leverages expert-authored long-text captions
            to enhance its medical language generation capabilities, enabling
            the production of semantically rich and clinically relevant image
            descriptions.

        \item \textbf{We introduce a novel data augmentation strategy} that
            transfers descriptive knowledge from expert-labeled samples to
            weakly labeled images via a similarity-based retrieval and re-captioning
            mechanism. This improves the diversity and semantic richness of training
            captions without incurring additional annotation cost.
    \end{itemize}

    Comprehensive experiments on a curated multi-disease oral dataset
    demonstrate that \textbf{OralGPT not only achieves competitive
    classification performance but also produces medically relevant, fluent image
    descriptions}. Our findings highlight the potential of vision-language
    systems in supporting clinical decision-making and advancing automated oral
    disease diagnosis.

    \section{Related Work}

    \subsection{Deep Learning and Vision-Language Models for Oral Disease}

    In recent years, deep learning has shown strong potential in the
    classification of oral mucosal diseases such as leukoplakia, oral lichen planus
    (OLP), and recurrent aphthous ulcers (RAU). Early methods relied on handcrafted
    features \cite{cheng2016computer,alrashdan2016oral}, while recent CNN-based
    approaches have achieved promising performance in lesion detection and
    classification
    \cite{keser2023deep,zhou2024deep,su2025deep,peng2024oral,adeoye2024deep}. Segmentation methods such as DeepLabV3+ have shown strong performance in medical image segmentation (e.g., Dice $ \approx $ 0.86–0.90 in skin lesion tasks) \cite{masood2024optimized}, while SegFormer achieved Dice  $ \approx $ 0.71 on oral mucosal lesion segmentation datasets \cite{zhang2024research}, and lightweight CNNs achieved 89–92\% accuracy for
    leukoplakia diagnosis \cite{ramesh2025artificial}. However, most existing
    works focus on classification and segmentation, lacking the ability to produce
    descriptive, free-text outputs.

    Vision-language models (VLMs), such as CLIP and Flamingo \cite{radford2021learning,alayrac2022flamingo},
    have enabled aligned image-text representation learning. Domain-specific extensions
    like BioViL \cite{boecking2022making}, LLaVA-Med \cite{li2023llava}, PMC-CLIP
    \cite{lin2023pmc}, and HuatuoGPT-Vision \cite{chen2024huatuogpt} have
    adapted this paradigm to medical imaging. In dermatology, SkinGPT-4 and
    PanDerm demonstrate multimodal reasoning across clinical and dermoscopic
    views \cite{zhou2023skingpt,yan2025multimodal}. ChatCAD further shows potential
    in clinical decision support \cite{wang2023chatcad}. However, applications of
    VLMs to oral photographic diagnosis remain rare, due to limited high-quality
    image-text datasets and complex domain-specific language demands.

    \subsection{Caption Generation and Multimodal Reasoning in Medicine}

    Caption generation for medical images is a long-standing challenge. Early
    efforts using radiology datasets such as MIMIC-CXR and ROCO focused on chest
    X-ray captioning \cite{boecking2022making,pelka2018radiology}. Recent works like
    MiniGPT-4 and Med-Flamingo improved grounded generation capabilities, yet
    often fall short in clinical factual accuracy
    \cite{zhu2023minigpt,moor2023med,sun2023evaluating,liu2025radiology}. In dentistry,
    recent models have analyzed periodontal bone loss and staging from
    radiographs \cite{jiang2022two,krois2019deep,chang2020deep}, but descriptive
    generation for oral photographs remains underexplored.

    Two-stage multimodal frameworks like LLaVA \cite{liu2023visual} first learn visual
    representations and then apply instruction tuning. Similar designs have
    shown cross-task advantages in medical visual question answering (Med-VQA \cite{he2020pathvqa})
    and clinical text summarization (ClinicalDocSum \cite{van2023clinical}). Our
    OralGPT framework follows this strategy: Stage 1 builds concept-aware visual
    features, while Stage 2 integrates expert captions to enhance generation fluency
    and clinical relevance.

    \subsection{Weak Supervision and Two-Stage Multimodal Learning}

    Label scarcity is a central challenge in medical image analysis. Semi-supervised
    methods like Noisy Student \cite{xie2020self} and FixMatch \cite{sohn2020fixmatch}
    use pseudo-labeling to exploit unlabeled data. In clinical imaging, contrastive
    approaches like ConVIRT align image-report pairs to improve representation learning
    \cite{zhang2022contrastive}. Inspired by these ideas, we propose similarity-guided
    caption propagation, which leverages both vision-language embedding
    similarity and textual relevance to transfer expert annotations to unlabeled
    oral images, enabling scalable fine-grained supervision in our setting.

    \section{Preliminary}

    To support our proposed OralGPT framework, we first present foundational concepts
    related to vision-language models, pseudo-labeling under limited supervision,
    and multimodal inference via prompts. We also formalize the key notations
    and embedding mechanisms adopted throughout this work.

    \subsection{Multimodal Vision-Language Embedding Space}

    Modern vision-language models (VLMs) are trained to map image and text inputs
    into a shared embedding space, enabling joint understanding and generation
    across modalities. Formally, given an image
    $x \in \mathbb{R}^{H \times W \times 3}$ and a corresponding caption or
    prompt $t \in \mathcal{T}$, the VLM encoders $\mathcal{F}_{\text{img}}$ and $\mathcal{F}
    _{\text{text}}$ produce fixed-dimensional embeddings:
    \[
        \mathbf{v}_{x}= \mathcal{F}_{\text{img}}(x), \quad \mathbf{v}_{t}= \mathcal{F}
        _{\text{text}}(t), \quad \mathbf{v}_{x}, \mathbf{v}_{t}\in \mathbb{R}^{d}
    \]

    where $d$ denotes the latent space dimensionality. The alignment objective
    during pretraining is typically contrastive, aiming to maximize the similarity
    between matched image-text pairs:

    \[
        \mathcal{L}_{\text{contrastive}}= -\log \frac{\exp(\text{sim}(\mathbf{v}_{x},
        \mathbf{v}_{t})/\tau)}{\sum_{t'}\exp(\text{sim}(\mathbf{v}_{x},
        \mathbf{v}_{t'})/\tau)}
    \]

    where $\text{sim}(\cdot,\cdot)$ is cosine similarity and $\tau$ is a
    learnable temperature.

    \subsection{CLIP-Based Feature Extraction and Similarity Retrieval}

    We utilize the CLIP-ViT model to compute image embeddings for retrieval-based
    pseudo-captioning. Let $\mathcal{D}_{\text{Partial}}= \{x_{q}\}$ be a set of
    query images (without captions), and $\mathcal{D}_{\text{Full}}= \{(x_{i}, c_{i}
    )\}$ be a labeled dataset with expert-written captions. For each $x_{q}$, we
    retrieve the top-$k$ most similar images in $\mathcal{D}_{\text{Full}}$:

    \[
        \mathcal{N}_{k}(x_{q}) = \arg\max_{x_i \in \mathcal{D}_{\text{Full}}}\text{sim}
        (\mathcal{F}_{\text{img}}(x_{q}), \mathcal{F}_{\text{img}}(x_{i})), \quad
        |\mathcal{N}_{k}| = k
    \]

    The corresponding captions $\{c_{i}\}_{i=1}^{k}$ are then used to construct
    a pseudo-caption $\hat{c}_{q}$ through filtering and paraphrasing.

    \subsection{Weak Supervision and Pseudo-Labeling Formalism}

    Let $\mathcal{D}_{\text{Full}}$ be the expert-labeled dataset containing $(x_{i}
    , y_{i}, c_{i})$ triplets, where $y_{i}$ is the disease label and $c_{i}$
    the natural language caption. Let $\mathcal{D}_{\text{Partial}}= \{(x_{j}, y_{j})
    \}$ be weakly labeled datasets
    with only image-class pairs.

    We define the pseudo-caption generation operator $\mathcal{P}: \mathbb{R}^{H
    \times W \times 3}\to \mathcal{T}$ such that:

    \[
        \hat{c}_{j}= \mathcal{P}(x_{j}; \mathcal{D}_{\text{Full}}), \quad \forall
        (x_{j}, y_{j}) \in \mathcal{D}_{\text{Partial}}
    \]

    The resulting pseudo-labeled dataset
    $\hat{\mathcal{D}}_{\text{Partial}}= \{(x_{j}, y_{j}, \hat{c}_{j})\}$ augments
    the training corpus for multimodal captioning.

\begin{figure}[htbp]
    \centering
    \includegraphics[width=1\linewidth]{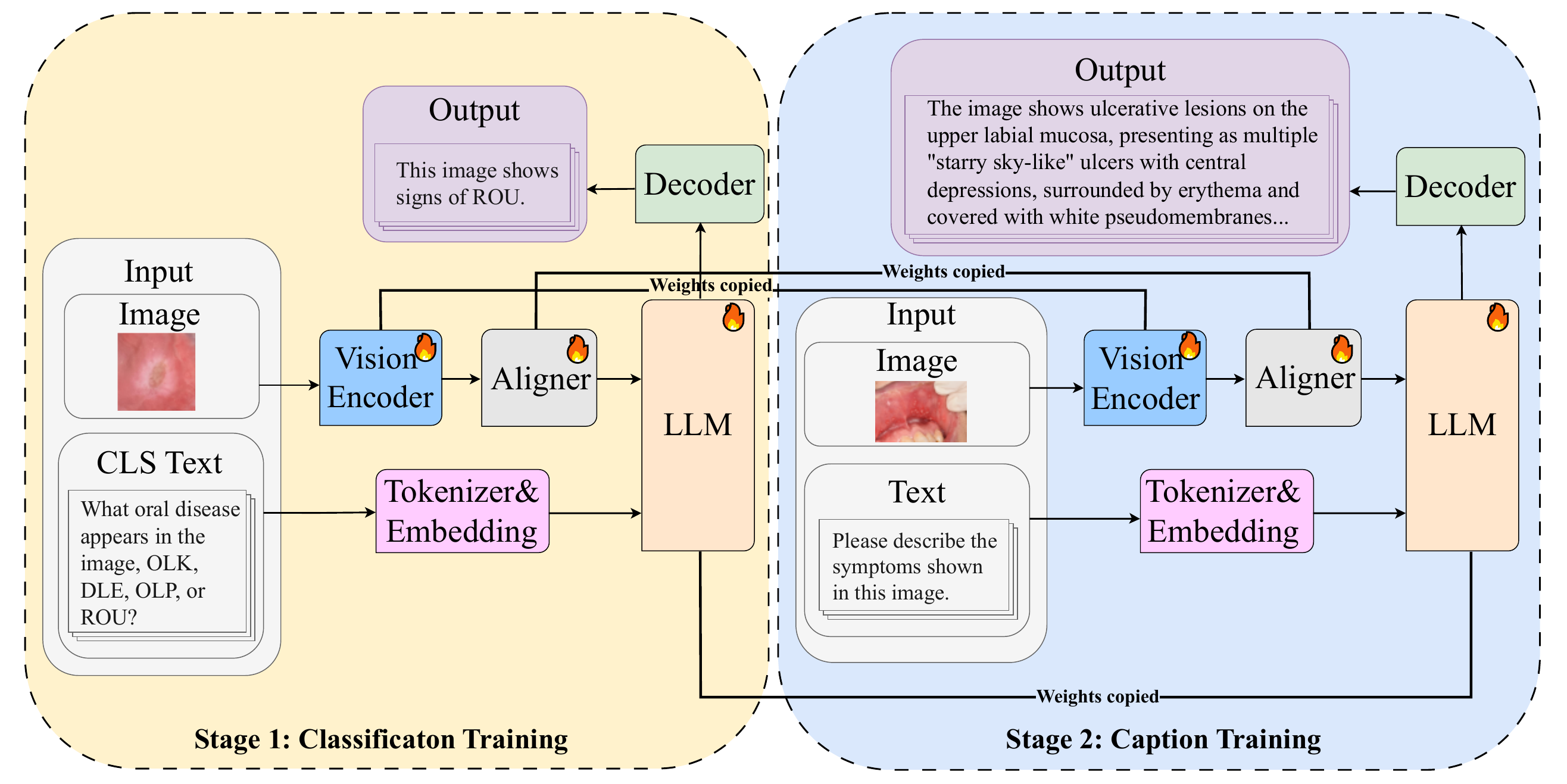}
    \caption{\textbf{Two-stage training pipeline of OralGPT.} Stage 1 (Classification Training): The model receives oral images and classification prompts, processes them through vision and text encoders, and outputs binary classification decisions for visual-textual alignment learning. Stage 2 (Caption Training): Inherits Stage 1 parameters, receives the same images with captioning prompts, and outputs detailed clinical descriptions.}
    \label{fig2}
\end{figure}

    \section{Methodology}

    Our proposed framework, \textbf{OralGPT}, adopts a two-stage vision-language
    modeling pipeline to address oral mucosal disease diagnosis and medical
    captioning under limited supervision. The framework leverages weak supervision,
    similarity-guided caption synthesis, and prompt-based multitask learning. In
    what follows, we describe the architecture, data handling, pseudo-caption strategy,
    and training details in a mathematically rigorous manner.


    \subsection{Overview of OralGPT Framework}

   OralGPT is built upon a multimodal foundation model (Qwen2.5-VL-7B-Instruct), which is capable of both visual-textual alignment and generative captioning. Leveraging this capacity, our framework is designed to facilitate understanding and generation tasks within the domain of oral medicine. Specifically, OralGPT comprises three interconnected components: a classification module \( \mathcal{G}_{\text{cls}}(x, t_{\text{cls}}) \) trained using prompt-based binary questions over multi-source datasets; a captioning module \( \mathcal{G}_{\text{cap}}(x, t_{\text{cap}}) \) supervised with both expert-written and pseudo-transferred natural language descriptions; and a retrieval-augmented pseudo-labeling module \( \mathcal{P}(x_q) \) that synthesizes descriptive captions for unlabeled or weakly labeled data by leveraging cross-modal similarity and filtering mechanisms.

Figure~\ref{fig2} illustrates the two-stage training pipeline of the proposed OralGPT framework. In \textbf{Stage 1 (Classification Training)}, the model learns to align visual and textual modalities through supervised classification prompts. Each training instance consists of an image and a corresponding prompt formulated as a binary question (e.g., ``What oral disease appears in the image: OLK, DLE, OLP, or ROU?''), where the image is processed by a vision encoder and the prompt by a tokenizer and embedding module. These two modalities are integrated via an alignment module and passed into a large language model (LLM), which outputs a binary decision. This stage primarily encourages the model to capture high-level semantic cues that associate image content with disease labels.

In \textbf{Stage 2 (Caption Training)}, the model is further trained to generate detailed free-text descriptions. The parameters from Stage 1 are reused to initialize all shared components, allowing for efficient knowledge transfer. During this phase, the input consists of the same image along with a captioning prompt (e.g., ``Please describe the symptoms shown in this image.''). The model is optimized to produce clinically coherent, fine-grained textual outputs that describe lesion characteristics, location, and visual patterns. A decoder module is used in both stages to generate output text from the fused multimodal representations.

This two-stage training strategy enables OralGPT to first develop a strong visual-language grounding through prompt-based classification, and then leverage this alignment for effective caption generation. Notably, by integrating retrieval-augmented pseudo-captioning, the model can be scaled to weakly labeled data without requiring full manual annotation, thereby enhancing generalization while minimizing annotation costs. Overall, OralGPT provides a unified and scalable solution for medical image understanding and reporting in oral healthcare.

\subsubsection{Prompt Augmentation}

To enhance language grounding, we employ prompt augmentation for both classification and captioning. Let \( \mathcal{P}_{\text{cls}} \) and \( \mathcal{P}_{\text{cap}} \) denote the respective prompt pools. For classification, each sample is paired with a binary question of the form:
\[
t_{\text{cls}}^{(y)} = \texttt{``Does this image show [disease $y$]?''}
\]
For captioning, we sample \( t_{\text{cap}} \sim \mathcal{P}_{\text{cap}} \), where \( |\mathcal{P}_{\text{cap}}| = 10 \). We use 3 classification prompts per image, and 6 prompts for each expert-captioned image during caption training.

\subsection{Similarity-Guided Pseudo-Caption Generation}
\label{sec:simtransfer}

To address the lack of textual descriptions in \( \mathcal{D}_{\text{Partial}} \), we propose a pseudo-caption generation module based on cross-modal retrieval. We first obtain normalized visual embeddings using a CLIP-ViT encoder \( f: \mathbb{R}^{H \times W \times 3} \to \mathbb{R}^d \), such that for any image \( x \in \mathcal{D}_{\text{Full}} \cup \mathcal{D}_{\text{Partial}} \), the embedding is:
\[
\mathbf{v}_x = \frac{f(x)}{\|f(x)\|}
\]
For a query image \( x_q \in \mathcal{D}_{\text{Partial}} \), we retrieve the top-\( k \) nearest neighbors from \( \mathcal{D}_{\text{Full}} \) using cosine similarity:
\[
\mathcal{N}_{k}(x_q) = \arg\max_{x_i \in \mathcal{D}_{\text{Full}}} \langle \mathbf{v}_{x_q}, \mathbf{v}_{x_i} \rangle
\]
The corresponding captions \( \{c_i\} \) are then filtered to remove redundancy and low-quality content. Specifically, we define:
\[
\hat{c}_q = \text{Filter}(\{c_i\}_{i=1}^k)
\]
where the \texttt{Filter} function removes (i) duplicate or near-duplicate sentences, (ii) captions containing fewer than three medical terms, and (iii) repetitive templates (e.g., ``white patch on mucosa''). Each resulting \( \hat{c}_q \) is then embedded into three randomly sampled prompts from \( \mathcal{P}_{\text{cap}} \) to form supervised training pairs for captioning.

\subsection{Training Objectives and Optimization}

Our model \( \mathcal{G}_\theta \) is jointly optimized across classification and captioning tasks using task-specific decoding objectives.

\paragraph{Classification Objective}  
Given input \( (x, y, t_{\text{cls}}) \), we model the probability \( p(y \mid x, t_{\text{cls}}) \) and optimize a binary cross-entropy loss over class-specific prompts:
\[
\mathcal{L}_{\text{cls}} = - \sum_{y} \left[ \mathbf{1}(y) \log p(y \mid x, t_{\text{cls}}) + (1 - \mathbf{1}(y)) \log (1 - p(y \mid x, t_{\text{cls}})) \right]
\]

\paragraph{Captioning Objective}  
Given a caption \( c = \{c_1, \ldots, c_n\} \) and input \( (x, t_{\text{cap}}) \), the decoder maximizes the likelihood of the next token:
\[
\mathcal{L}_{\text{cap}} = - \sum_{t=1}^{n} \log p(c_t \mid x, t_{\text{cap}}, c_{<t})
\]

\paragraph{Optimization Details}  
Model training is formulated as a two-stage optimization process. In \textbf{Stage 1}, the parameters of the vision-language model are fine-tuned on the classification objective over both fully and weakly labeled datasets. Input images are rescaled based on source characteristics to ensure consistent representation, and prompt-based supervision is employed to guide multimodal alignment. Optimization is performed using a gradient-based method with adaptive weight decay, accompanied by warmup scheduling and gradient accumulation to stabilize updates under constrained batch regimes. In \textbf{Stage 2}, the model is initialized with parameters from Stage 1 and further optimized with the captioning objective on both human-annotated and pseudo-labeled data. High-resolution image inputs are preserved to retain fine-grained spatial information, and no augmentation is applied to maintain semantic fidelity in generative tasks.

    \section{Experiment}
    \subsection{Datasets and Data Preparation}
    \subsubsection{Dataset Overview}
    To support the development and evaluation of the proposed \textit{OralGPT} system,
    we curated a multi-source dataset covering four visually similar and
    clinically relevant oral diseases: \textbf{Oral Leukoplakia (OLK)}, \textbf{Oral
    Lichen Planus (OLP)}, \textbf{Recurrent Oral Aphthous Ulcer (ROU)}, and \textbf{Discoid
    Lupus Erythematosus (DLE)}.

    The data collection for this study includes clinical images of patients who
    visited the Department of Oral Mucosal Diseases at the Hospital of
    Stomatological Xi'an Jiaotong University between January 2024 and April 2025,
    including oral lichen planus, chronic discoid lupus erythematosus, recurrent
    aphthous ulcer, and oral leukoplakia. All included cases were confirmed through
    clinical or histopathological diagnosis. Two senior oral mucosal doctors
    provide professional descriptions of the images and annotate the lesion
    characteristics of each image.

As detailed in Section 4.2, our dataset comprises three subsets: the fully-annotated
D\textsubscript{Full}, the weakly-labeled private D\textsubscript{Partial}, and
the weakly-labeled public D\textsubscript{Public}. The specific image counts for
each disease across these subsets are provided in the tables below.

    \vspace{0.5em}
    \noindent
    \textbf{D\textsubscript{Full}} is a private dataset labeled by professional
    oral clinicians with both disease categories and natural language symptom
    descriptions. It serves as the primary supervised training and evaluation source
    for both classification and captioning tasks.

    We performed a stratified 70\%/30\% train-test split to create D\textsubscript{Full\_train} and D\textsubscript{Full\_test}.

    \vspace{0.5em}
    \noindent
    \textbf{D\textsubscript{Partial}} is a private weakly-labeled dataset containing only disease class annotations, without expert textual descriptions. It is used to supplement the classification task and is enhanced with pseudo-descriptions during Stage 2 using image similarity-guided caption transfer. The number of images in D\textsubscript{Full} and D\textsubscript{Partial} is shown in Table~\ref{tab:pro_data_summary}.

    \begin{table}[H]
        \centering
        \caption{Image Counts in private datasets (D\textsubscript{Full} and D\textsubscript{Partial})
        }
        \begin{tabular}{lcccc}
            \toprule \multirow{2}{*}{\textbf{Disease}} & \multicolumn{3}{c}{\textbf{D\textsubscript{Full}}} & \multirow{2}{*}{\textbf{D\textsubscript{Partial}}}     \\
            \cmidrule{2-4}                             & \textbf{D\textsubscript{Full}}                     & \textbf{D\textsubscript{Full\_train}} & \textbf{D\textsubscript{Full\_test}} &              \\
            \midrule OLK                               & 116                                                & 81                                    & 35                                   & 79           \\
            OLP                                        & 196                                                & 137                                   & 59                                   & 514          \\
            ROU                                        & 83                                                 & 58                                    & 25                                   & 27           \\
            DLE                                        & 85                                                 & 59                                    & 26                                   & 39           \\
            \midrule \textbf{Total}                    & \textbf{480}                                       & \textbf{335}                          & \textbf{145}                         & \textbf{659} \\
            \bottomrule
        \end{tabular}
        \label{tab:pro_data_summary}
    \end{table}

    \vspace{0.5em}
    \noindent
    \textbf{D\textsubscript{Public}} is a public weakly-labeled dataset aggregated
    from multiple open-access oral image sources. Each image is annotated only
    with a disease category. Due to heterogeneous quality and potential label noise,
    a strict cleaning and filtering process is applied (see Section~\ref{sec:cleaning}).The number of images in D\textsubscript{Public} is shown in Table~\ref{tab:Dpublic}. In this table, a dash (-) indicates that the corresponding data source does not contain images for that disease category.

    \begin{table}[H]
        \centering
        \caption{Image Counts in D\textsubscript{Public} Before and After
        Cleaning }
        \resizebox{1\textwidth}{!}{ 
        \begin{tabular}{lccccc}
            \toprule \textbf{Source}              & \textbf{OLK} & \textbf{OLP} & \textbf{ROU} & \textbf{DLE} & \textbf{Total} \\
            \midrule leukoplakia-8woh8(Roboflow)\tablefootnote{leukoplakia-8woh8 https://universe.roboflow.com/makhluk/leukoplakia-8woh8}  & 232          & \  -         & \  -         & \  -         & 232            \\
            leukoplakia(Roboflow)\tablefootnote{leukoplakia https://universe.roboflow.com/oral-cancer-emoot/leukoplakia}                 & 74           & \ -          & \  -         & \  -         & 74             \\
            OralLeukoplakia(Roboflow)  \tablefootnote{OralLeukoplakia https://universe.roboflow.com/oral-dquwv/oralleukoplakia}           & 418          & \  -         & \  -         & \  -         & 418            \\
            ulcer\_detection-12cr0(Roboflow) \tablefootnote{ulcer\_detection-12cr0 https://universe.roboflow.com/aadith/ulcer\_detection-12cr0}     & \  -         & \  -         & 382          & \  -         & 382            \\
            Oral Diseases(Kaggle) \tablefootnote{Oral Diseases https://www.kaggle.com/datasets/salmansajid05/oral-diseases}                & \  -         & \  -         & 265          & \  -         & 265            \\
            Mouth and Oral Diseases (MOD)(Kaggle) \tablefootnote{Mouth and Oral Diseases https://www.kaggle.com/datasets/javedrashid/mouth-and-oral-diseases-mod?select=Training}& \  -         & 74           & \  -         & \  -         & 74             \\
            lichen\_planus(Roboflow) \tablefootnote{lichen\_planus https://universe.roboflow.com/lichenplanus}             & \  -         & 25           & \  -         & \  -         & 25             \\
            \midrule \textbf{Total(Raw)}          & 724          & 99           & 647          & \ -          & 1470           \\
            \textbf{Total(Cleaned)}               & 624          & 60           & 596          & \  -         & 1280           \\
            \bottomrule
        \end{tabular}
        }
        \label{tab:Dpublic}
    \end{table}
    
    Unless otherwise stated, D\textsubscript{Full} in subsequent sections refers
    to the training split (D\textsubscript{Full\_train}), and D\textsubscript{test}
    refers to (D\textsubscript{Full\_test}).

    \subsubsection{Data Cleaning and Augmentation for Classification}
    \label{sec:cleaning}

    For the classification task (Stage 1), all datasets underwent a rigorous preprocessing pipeline involving data cleaning, image processing, and data augmentation.

    \paragraph{Data Cleaning} A multi-step cleaning process was applied to ensure data quality, which was particularly critical for the heterogeneous D\textsubscript{Public} dataset. First, all images were manually reviewed by domain experts to identify and remove samples with incorrect or ambiguous disease labels. Second, we filtered out low-quality images characterized by issues such as severe blurriness, poor lighting, out-of-focus areas, or significant visual obstructions that obscured the lesion. Finally, duplicate and near-duplicate images were programmatically identified and removed to prevent data redundancy. This comprehensive filtering is reflected in the reduced image counts for D\textsubscript{Public} shown in Table~\ref{tab:Dpublic}.

    \paragraph{Image Processing and Augmentation} Following cleaning, images were processed for model input. Images from the private datasets (D\textsubscript{Full}, D\textsubscript{Partial}) were cropped around the lesion and then resized to $560 \times 560$ pixels. In contrast, images from D\textsubscript{Public} were resized to $224 \times 224$ pixels. Subsequently, we applied the data augmentation techniques as detailed in Section~4.2. This included both image-level transformations (e.g., random flips, color jitter) and textual prompt augmentation, where three distinct prompts were sampled per image to enhance the model's robustness to varied textual queries.

    \subsubsection{Data Configuration for Caption Generation}

    For the caption generation task (Stage 2), the data was configured
    differently to optimize for descriptive text generation:

    The training process exclusively relied on private datasets: D\textsubscript{Full}
    (containing expert-authored descriptions) and D\textsubscript{Partial} (augmented
    with pseudo-captions via the similarity-guided strategy in Section~\ref{sec:simtransfer}),
    while excluding the D\textsubscript{Public} dataset. All images were processed
    at their native full resolution (4000×6000 pixels) to preserve contextual details,
    with no lesion-centered cropping or image-level augmentations applied. For prompt
    augmentation, each image in D\textsubscript{Full} and D\textsubscript{Partial}
    was randomly assigned 6 and 3 textual prompts respectively, following the method
    outlined in Section~4.2.2.

    This configuration aimed to leverage the foundational visual understanding
    developed in Stage 1, while providing the model with rich, unaltered visual
    input and diverse textual cues for generating high-quality medical captions.

    \subsection{Experimental Settings and Variants}
    \label{sec:exp_settings_variants}

    \paragraph{General Settings}
    We adopt Qwen2.5-VL-7B-Instruct as the backbone model for both stages. Fine-tuning
    is performed using the Swift framework with LoRA (rank=16, alpha=16) applied
    to all linear layers. Training uses the AdamW optimizer with a learning rate
    of 1e-4 and warmup ratio of 0.05. All experiments are conducted on 4 NVIDIA RTX
    4090 GPUs with bfloat16 precision.

    \paragraph{Configurations for Stage 1 (Classification)}
    Input images are resized to 560\,$\times$\,560 (private datasets) or 224\,$\times$\,224
    (public datasets). The model is trained for three epochs with a batch size of
    1 per GPU and gradient accumulation steps of 16. We compare training
    configurations to evaluate the impact of weakly labeled data and multi-source
    integration as shown in Table~\ref{tab:classification_configs}.


\begin{table}[htbp]
\footnotesize
\centering
\caption{Training configurations for classification experiments (Stage 1)}
\label{tab:classification_configs}
\renewcommand{\arraystretch}{1.4}
\begin{tabularx}{\textwidth}{
  >{\raggedright\arraybackslash}p{4.5cm}
  >{\raggedright\arraybackslash}X
}
\rowcolor{gray!15}
\toprule
 \multicolumn{1}{c}{\textbf{Model Variant}} & \multicolumn{1}{c}{\textbf{Description}} \\
\midrule
\rowcolor{blue!8}
\textbf{Zeroshot} &
Directly evaluates the pretrained Qwen2.5-VL-7B model on binary classification prompts without any fine-tuning. \\[8pt]
\hline
\rowcolor{gray!5}
\textbf{D\textsubscript{Full} Only} \newline
{\small \textcolor{gray}{(Baseline)}} &
Trains the model on the private, fully annotated dataset D\textsubscript{Full}. \\[8pt]
\hline
\rowcolor{blue!8}
\textbf{D\textsubscript{Full} + D\textsubscript{Partial}} &
Adds weakly labeled private data (D\textsubscript{Partial}) to the baseline training set. \\[8pt]
\hline
\rowcolor{gray!5}
\textbf{D\textsubscript{Full} + D\textsubscript{Public}} &
Uses expert-annotated private data combined with public weakly labeled data. \\[8pt]
\hline
\rowcolor{green!12}
\textbf{D\textsubscript{Full} + D\textsubscript{Partial} + D\textsubscript{Public}} \newline
{\small \textcolor{blue}{\textbf{(Stage 1 final model)}}} &
Incorporates both private and public weakly labeled data into training. This configuration represents our comprehensive final Stage 1 model. \\[8pt]
\hline
\rowcolor{blue!8}
\textbf{HuatuoGPT-Vision} \newline
{\small \textcolor{gray}{\cite{chen2024huatuogpt}}} &
Evaluates the pretrained HuatuoGPT-Vision model on binary classification prompts using both image and text inputs, without task-specific fine-tuning. \\[8pt]
\bottomrule
\end{tabularx}
\end{table}

\vspace{1cm}

    \paragraph{Configurations for Stage 2 (Caption Generation)}
    The model is initialized from the Stage~1 checkpoint (unless otherwise specified)
    and trained using original full-resolution images (4000\,$\times$\,6000). No
    image augmentation is applied at this stage. Prompt augmentation for
    captioning is applied as described in Section~5.1.3. We compare four
    training configurations as shown in Table~\ref{tab:captioning_configs}


\begin{table}[htbp]
\footnotesize
\centering
\caption{Training configurations for captioning experiments (Stage 2)}
\label{tab:captioning_configs}
\renewcommand{\arraystretch}{1.4}
\begin{tabularx}{\textwidth}{
  >{\raggedright\arraybackslash}p{4.5cm}
  >{\raggedright\arraybackslash}X
}
\rowcolor{gray!15}
\toprule
\multicolumn{1}{c}{\textbf{Model Variant}} & \multicolumn{1}{c}{\textbf{Description}} \\
\midrule
\rowcolor{blue!8}
\textbf{Stage~1 + D\textsubscript{Full}} \newline
{\small \textcolor{gray}{(Baseline)}} &
Caption generation is fine-tuned on top of the Stage~1 classification model using expert-annotated image-text pairs from D\textsubscript{Full} only. \\[8pt]
\hline
\rowcolor{green!12}
\textbf{Stage~1 + D\textsubscript{Full} + D\textsubscript{Partial}} \newline
{\small \textcolor{blue}{\textbf{(Stage 2 final model)}}} &
Adds pseudo-captions generated via the similarity-guided method from D\textsubscript{Partial} (see Section~\ref{sec:simtransfer}). Both D\textsubscript{Full} and D\textsubscript{Partial} samples are used for caption training. \\[8pt]
\hline
\rowcolor{blue!8}
\textbf{Direct D\textsubscript{Full}} \newline
{\small \textcolor{gray}{(Without Stage~1)}} &
Trains captioning directly on the original Qwen2.5-VL-7B model using D\textsubscript{Full}, without pre-training of Stage~1. \\[8pt]
\hline
\rowcolor{gray!5}
\textbf{Direct D\textsubscript{Full} + D\textsubscript{Partial}} \newline
{\small \textcolor{gray}{(Without Stage~1)}} &
Same as above, but also includes pseudo-captioned D\textsubscript{Partial}. \\[8pt]
\bottomrule
\end{tabularx}
\end{table}

\vspace{1cm}

\subsection{Evaluation Metrics}
\label{sec:eval_metrics}
To assess the performance of our model across both classification and captioning tasks, we adopt tailored evaluation protocols and metrics suited to the specific output format and clinical relevance of each task.

\paragraph{Classification Task Evaluation Metrics}
While the model is trained using a four-class classification objective, the evaluation on D\textsubscript{full\_test} adopts a clinically meaningful binary evaluation framework. For each test sample and target disease \textit{d} $\in$ \{1,...,k\} (where k=4 in our case), we formulate a binary decision task: ``Does this image show disease \textit{d}?". The model generates a probability score $p_d \in [0,1]$ for each disease class, with binary predictions $\hat{y}_d$ obtained by thresholding at 0.5:

\begin{equation*}
    \hat{y}_d = \begin{cases}
        1 & \text{if } p_d \geq 0.5 \\
        0 & \text{otherwise}
    \end{cases}
\end{equation*}

Let $\text{TP}_d$, $\text{FP}_d$, $\text{TN}_d$, and $\text{FN}_d$ denote respectively the true positives, false positives, true negatives, and false negatives for disease \textit{d}. We evaluate performance using the following metrics, computed per disease and macro-averaged across all diseases:

\begin{itemize}
    \item \textbf{Accuracy}: The overall proportion of correct predictions across all samples, measuring how often the model makes correct predictions regardless of class:
    \begin{equation*}
        \text{Acc}_d = \frac{\text{TP}_d + \text{TN}_d}{\text{TP}_d + \text{FP}_d + \text{TN}_d + \text{FN}_d}
    \end{equation*}
    While accuracy provides an intuitive overall measure, it can be misleading in imbalanced datasets where one class dominates.
    
    \item \textbf{Precision}: Quantifies the reliability of positive predictions by measuring the proportion of positive predictions that are actually correct:
    \begin{equation*}
        P_d = \frac{\text{TP}_d}{\text{TP}_d + \text{FP}_d}
    \end{equation*}
    High precision minimizes false positive errors, which is crucial in clinical settings to reduce false alarms and prevent unnecessary follow-up procedures.
    
    \item \textbf{Recall}: Measures the model's ability to correctly identify all actual positive cases within the dataset:
    \begin{equation*}
        R_d = \frac{\text{TP}_d}{\text{TP}_d + \text{FN}_d}
    \end{equation*}
    High recall is critical for early detection and preventing missed diagnoses that could lead to delayed treatment or adverse patient outcomes.
    
    \item \textbf{F1-Score\cite{vanrijsbergen1979}}: The harmonic mean of precision and recall, providing a balanced measure that accounts for both false positive and false negative errors:
    \begin{equation*}
        \text{F1}_d = 2 \cdot \frac{P_d \cdot R_d}{P_d + R_d} = \frac{2 \cdot \text{TP}_d}{2 \cdot \text{TP}_d + \text{FP}_d + \text{FN}_d}
    \end{equation*}
    The F1-score is particularly valuable when dealing with imbalanced datasets or when both precision and recall are equally important for clinical decision-making.
\end{itemize}

The macro-averaged metrics are computed as:

\begin{equation*}
    \overline{\text{Metric}} = \frac{1}{k}\sum_{d=1}^{k} \text{Metric}_d
\end{equation*}

where $k$ is the number of diseases. This approach ensures equal weighting of all conditions regardless of their prevalence in the test set, providing a balanced assessment of the model's performance across all oral mucosal conditions.

\paragraph{Caption Generation Task Evaluation Metrics}
The caption generation performance is evaluated using both lexical overlap metrics and clinical semantic assessment.

\subparagraph{Lexical Overlap Metrics}
To quantify textual similarity between model-generated captions and expert reference descriptions, we employ established automatic evaluation metrics that assess different linguistic aspects of caption quality:

\begin{itemize}
    \item \textbf{BLEU-n} \cite{papineni2002bleu}: Measures n-gram precision between generated and reference captions, evaluating fluency and adequacy at multiple granularity levels:
    \begin{equation*}
        \text{BLEU-n} = BP \cdot \exp\left(\sum_{k=1}^n w_k \log p_k\right)
    \end{equation*}
    where $p_k$ represents the k-gram precision, $w_k = 1/n$ are uniform weights across different n-gram orders, and $BP$ is the brevity penalty that compensates for overly short generations:
    \begin{equation*}
        BP = \begin{cases}
            1 & \text{if } l_c > l_r \\
            e^{1-l_r/l_c} & \text{otherwise}
        \end{cases}
    \end{equation*}
    where $l_c$ and $l_r$ denote the lengths of candidate and reference captions, respectively. BLEU-4 is particularly valuable for medical caption evaluation as it captures multi-word clinical terminology accuracy.
    
    \item \textbf{METEOR} \cite{banerjee2005meteor}: Evaluates caption quality through word-level alignment that incorporates exact matches, stemming, and synonymy. Unlike BLEU's precision-focused approach, METEOR provides a balanced assessment by considering both precision and recall:
    \begin{equation*}
        \text{METEOR} = (1 - \gamma) \cdot \frac{10PR}{R+9P} + \gamma \cdot \text{Penalty}
    \end{equation*}
    where $P$ and $R$ represent precision and recall of aligned unigrams, and $\gamma$ controls the fragmentation penalty for non-contiguous word matches. METEOR's ability to recognize synonymous expressions makes it suitable for medical texts where clinical terms may have multiple valid formulations.
    
    \item \textbf{ROUGE} \cite{lin2004rouge}: Assesses content coverage and recall-oriented quality through multiple complementary variants:
    \begin{itemize}
        \item ROUGE-1 measures unigram overlap to capture content similarity:
        \begin{equation*}
            R_1 = \frac{\sum_{w\in r} \min(\text{count}(w,c), \text{count}(w,r))}{\sum_{w\in r} \text{count}(w,r)}
        \end{equation*}
        \item ROUGE-L evaluates structural coherence through longest common subsequence:
        \begin{equation*}
            R_L = \frac{\text{LCS}(c,r)}{l_r}
        \end{equation*}
        where $\text{LCS}(c,r)$ represents the longest common subsequence between candidate $c$ and reference $r$, and $l_r$ is the reference length.
    \end{itemize}
    ROUGE metrics are particularly informative for evaluating whether key clinical observations (lesion morphology, anatomical location, color characteristics) are preserved in generated descriptions.
\end{itemize}

While these lexical metrics provide objective measures of textual similarity, they primarily capture surface-level overlap and may not fully reflect the clinical accuracy or semantic appropriateness of generated medical descriptions.

\subparagraph{Clinical Semantic Evaluation}
To assess the clinical quality of model-generated captions beyond surface-level lexical similarity, we utilize the DeepSeek-V3 large language model as an automatic evaluator. Given a reference label and the model’s output, DeepSeek-V3 produces scores along three dimensions: completeness (i.e., the extent to which key clinical findings are covered), professionalism (i.e., the appropriate use of medical terminology), and correctness (i.e., the clinical accuracy of the statements). The final score is computed as the average of these three aspects, offering a holistic evaluation of caption quality that complements traditional lexical matching metrics.

    \subsection{Results}
    \label{sec:results}

    This section presents the experimental results for the classification task (Stage
    1), the caption generation task (Stage 2), and the unified evaluation of the
    final model. All configurations are as defined in Section~\ref{sec:exp_settings_variants}.

\paragraph{Stage 1: Classification Performance}
Table~\ref{tab:stage1-classification} summarizes the binary classification performance on D\textsubscript{test}. The zeroshot model performed poorly, underscoring the need for domain-specific fine-tuning. Our Stage 1 final model (D\textsubscript{Full} + D\textsubscript{Partial} + D\textsubscript{Public}) achieved the best overall balance, with the highest accuracy (77.24\%), precision (83\%), and F1-score (75.06\%), demonstrating the clear benefit of multi-source data integration over using D\textsubscript{Full} alone. For comparison, while the general medical HuatuoGPT-Vision achieved a very high recall (90.80\%), its precision was low (53.23\%), indicating a tendency to predict positive for most cases, which limits its practical clinical utility. This contrast highlights the effectiveness of our domain-specific, multi-source training approach.

    \begin{table}[H]
        \centering
        \caption{Binary classification results (per disease averaged). All models
        are evaluated on D\textsubscript{test} using binary prompts.}
        \label{tab:stage1-classification} \resizebox{1\textwidth}{!}{
        \begin{tabular}{lcccc}
            \toprule \textbf{Model Variant}                                                    & \textbf{Accuracy(\%)} & \textbf{Precision(\%)} & \textbf{Recall(\%)} & \textbf{F1-score(\%)} \\
            \midrule Zeroshot                                                                  & 54.25                 & 53.41                  & 66.67               & 59.3                  \\
            D\textsubscript{Full} Only (Baseline)                                              & 66.67                 & 71.39                  & 55.63               & 62.53                 \\
            D\textsubscript{Full} + D\textsubscript{Partial}                                   & 74.02                 & 75.8                   & \textbf{70.57}      & 73.1                  \\
            D\textsubscript{Full} + D\textsubscript{Public}                                    & 63.56                 & 64.05                  & 61.84               & 62.92                 \\
            HuatuoGPT-Vision                                                                          & 55.51                 & 53.23                  & 90.80               & 67.11                 \\
            D\textsubscript{Full} + D\textsubscript{Partial} + D\textsubscript{Public} (Final) & \textbf{77.24}        & \textbf{83}            & 68.5                & \textbf{75.06}        \\

            \bottomrule
        \end{tabular}
        }
    \end{table}

    \paragraph{Stage 2: Caption Generation Performance}
    Tables~\ref{tab:stage2-caption-lexical} and
    \ref{tab:stage2-caption-semantic} present the captioning model evaluations
    on D\textsubscript{test}. The Stage 2 final model (initialized from Stage 1,
    trained on D\textsubscript{Full} + D\textsubscript{Partial}) consistently
    outperformed other configurations across all lexical metrics (Table~\ref{tab:stage2-caption-lexical})
    and DeepSeek-V3 semantic metrics (Table~\ref{tab:stage2-caption-semantic}). Notably,
    it achieved the highest DeepSeek-V3 average score (6.36), indicating superior
    completeness, professionalism, and correctness. These results validate the
    effectiveness of similarity-guided pseudo-captioning and classification-aware
    pretraining for generating high-quality medical descriptions.

    \begin{table}[H]
        \centering
        \caption{Evaluation of captioning models on D\textsubscript{test} using
        lexical overlap metrics.}
        \label{tab:stage2-caption-lexical} \resizebox{1\textwidth}{!}{
        \begin{tabular}{ l@{\hskip 4pt} c@{\hskip 4pt} c@{\hskip 4pt} c@{\hskip 4pt}
        c@{\hskip 4pt} c }
            \toprule \textbf{Model}                                  & \textbf{BLEU-1} & \textbf{BLEU-4} & \textbf{METEOR} & \textbf{ROUGE-1} & \textbf{ROUGE-L} \\
            \midrule 
            Direct D\textsubscript{Full}                             & 0.5687          & 0.3356          & 0.5846          & 0.6415           & 0.5997           \\
            Direct D\textsubscript{Full} + D\textsubscript{Partial}  & 0.5755          & 0.3582          & 0.5881          & 0.6435           & 0.6038           \\            
            Stage~1 + D\textsubscript{Full} (Baseline)      & 0.5699          & 0.3400            & 0.5754          & 0.6439           & 0.6011           \\
            Stage~1 + D\textsubscript{Full} + D\textsubscript{Partial} (Final) & \textbf{0.5827} & \textbf{0.3736} & \textbf{0.5953} & \textbf{0.6525}  & \textbf{0.6129}  \\
            \bottomrule 
        \end{tabular}
        }
    \end{table}

    \begin{table}[H]
        \centering
        \caption{Evaluation of captioning models on D\textsubscript{test} using
        DeepSeek-V3 semantic scoring.}
        \label{tab:stage2-caption-semantic} \resizebox{1\textwidth}{!}{
        \begin{tabular}{ l@{\hskip 4pt} c@{\hskip 4pt} c@{\hskip 4pt} c@{\hskip 4pt}
        c }
            \toprule \textbf{Model}                                  & \textbf{Completeness} & \textbf{Profesionalism} & \textbf{Correctness} & \textbf{Avg.} \\
            \midrule 
            Direct D\textsubscript{Full}                             & 5.52                  & 5.89                    & 4.54                 & 5.32          \\
            Direct D\textsubscript{Full} + D\textsubscript{Partial}  & 6.28                  & 6.73                    & 5.60                 & 6.20           \\
            Stage~1 + D\textsubscript{Full} (Baseline)      & 5.88             & 6.26                    & 5.02                 & 5.72          \\
            Stage~1 + D\textsubscript{Full} + D\textsubscript{Partial} (Final) & \textbf{6.51}         & \textbf{6.89}           & \textbf{5.67}        & \textbf{6.36} \\
            \bottomrule
        \end{tabular}
        }
    \end{table}

    \paragraph{Unified Evaluation}

    To assess the impact of caption-focused training on classification capabilities, the Stage 2 final model was re-evaluated on the D\textsubscript{test} classification task. As shown in Table~\ref{tab:unified evaluation}, the Stage 2 model (F1-score: 78.13\%) not only maintained but slightly improved classification performance compared to the Stage 1 final model (F1-score: 75.06\%). These results indicate that supervision from caption generation enhances the model's semantic comprehension, thereby improving its classification accuracy.

    \begin{table}[H]
        \centering
        \caption{Binary classification accuracy on D\textsubscript{test} before
        and after caption training.}
        \label{tab:unified evaluation} \resizebox{1\textwidth}{!}{
        \begin{tabular}{lcccc}
            \toprule \textbf{Model}      & \textbf{Accuracy(\%)} & \textbf{Precision(\%)} & \textbf{Recall(\%)} & \textbf{F1-score(\%)} \\
            \midrule Stage~1 Final Model & 77.24                 & \textbf{83.00}            & 68.50                & 75.06                 \\
            Stage~2 Final Model          & \textbf{77.93}        & 77.43                  & \textbf{78.85}      & \textbf{78.13}        \\
            \bottomrule
        \end{tabular}
        }
    \end{table}

\begin{figure}[htbp]
    \centering
    \includegraphics[width=0.6\linewidth]{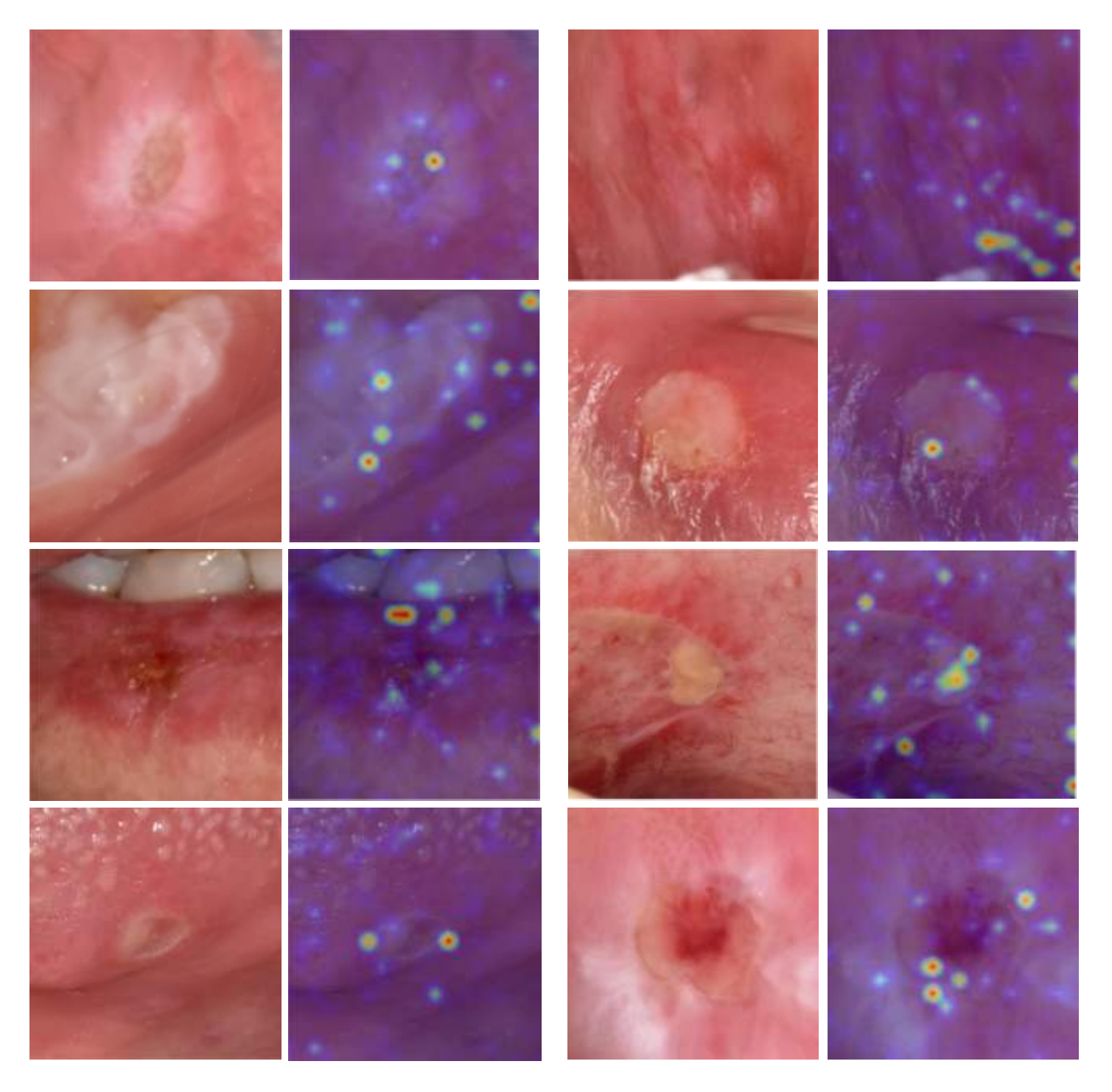}
    \caption{\textbf{Visualization of attention heatmaps for the classification task.} Each row displays a representative sample, where the left image shows the original oral lesion and the right image overlays the aggregated attention heatmap. The heatmaps are computed by averaging the attention weights across all 28 heads of the vision encoder.}
    \label{fig:attention_heatmap}
\end{figure}

    \paragraph{Results Visualization: Attention Mechanism}
To better interpret the model’s visual reasoning during disease classification, we present a series of attention visualizations in Figure~\ref{fig:attention_heatmap}. Each row displays a representative sample, where the left image shows the original oral lesion and the right image overlays the aggregated attention heatmap onto the same image. The heatmaps are computed by averaging the attention weights across all 28 heads of the vision encoder at the time of generating the first token in response to the classification prompt. These visualizations highlight the regions most influential to the model’s initial prediction, offering insight into which pathological features the model relies on for decision making.



\begin{figure}[htbp]
    \centering
    \includegraphics[width=0.8\linewidth]{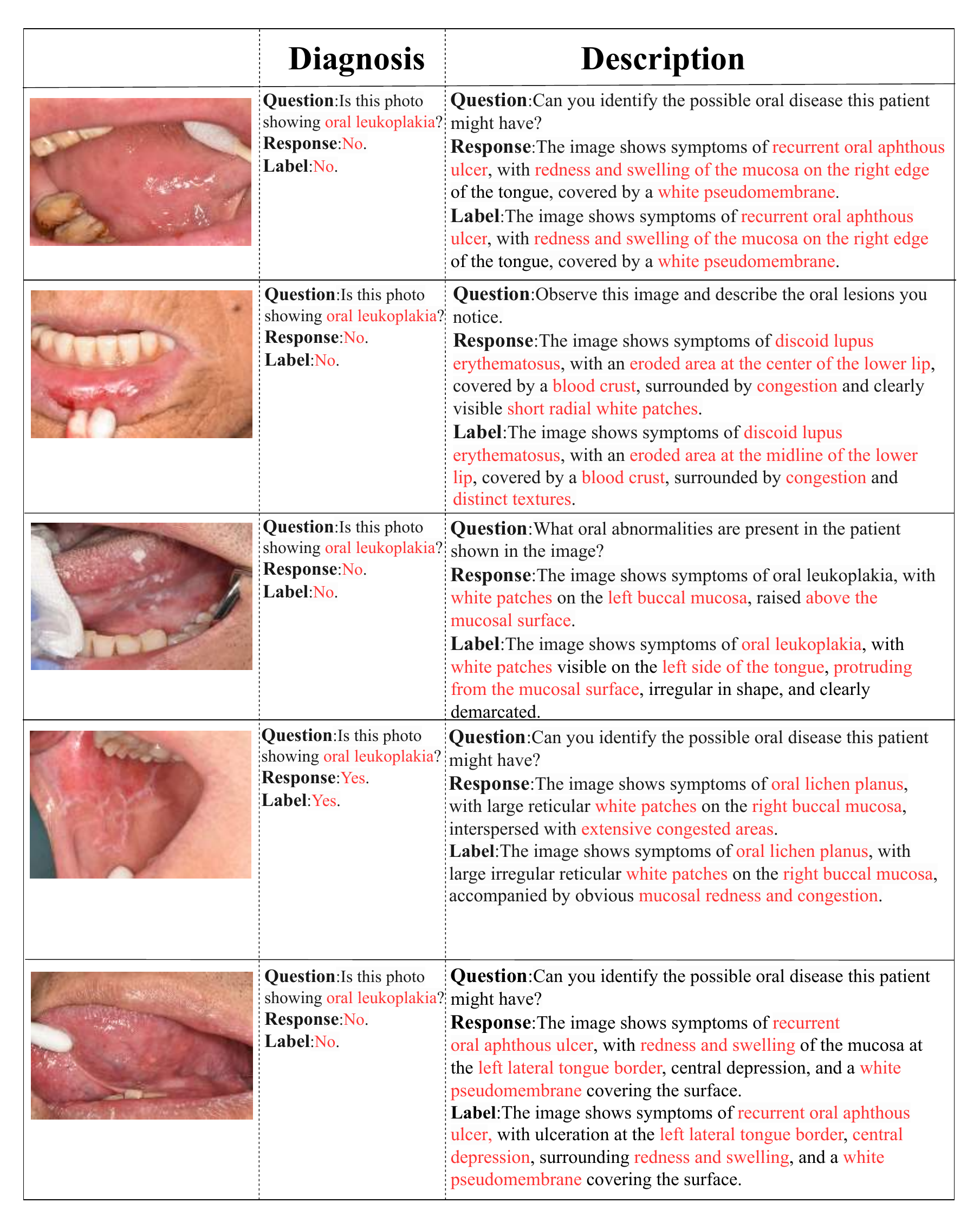}
    \caption{\textbf{Example of Model Prediction Results.} Each row shows one input image with three columns: the input photo, diagnosis response, and description response. The diagnosis column answers questions about possible diseases, while the description column provides detailed lesion descriptions. Reference labels are provided below each response, with key pathological terms highlighted in red. During training, we used Chinese data. For demonstration purposes, the content here has been translated into English.}
    \label{fig:result}
\end{figure}

To illustrate the differences in model behavior between the two stages, Figure~\ref{fig:result} presents representative examples of model predictions. Each row corresponds to one input image, with three columns showing the input photo, the diagnosis response (Stage 1), and the description response (Stage 2), respectively. In Stage 1, the model answers a yes/no question regarding the presence of oral leukoplakia. In Stage 2, the model responds to a more open-ended query, describing the observed lesion or identifying the likely disease. The reference label is provided below each response. 

We highlight in red the key pathological terms that are semantically consistent with the reference label. These examples demonstrate that Stage 2 responses tend to contain richer clinical details and more accurate lesion descriptors, even in cases where Stage 1 predictions are incorrect. This suggests that open-ended captioning facilitates more interpretable and informative outputs in clinical decision support contexts.


    \paragraph{Implication}
    The comprehensive experimental results strongly validate the OralGPT framework's
    efficacy. The Stage 1 classification results (Figure~\ref{Classification
    result}) underscore the critical role of multi-source data integration; combining
    D\textsubscript{Full} with weakly-labeled D\textsubscript{Partial} and D\textsubscript{Public}
    yielded superior diagnostic accuracy (77.24\% Acc, 75.06\% F1) compared to
    using D\textsubscript{Full} alone or other combinations, demonstrating that diverse,
    albeit less annotated, data significantly enriches the model's understanding
    of disease manifestations.This approach effectively mitigates the data scarcity
    problem common in specialized medical domains.

    \begin{figure}[htbp]
        \centering
        \includegraphics[width=1\linewidth]{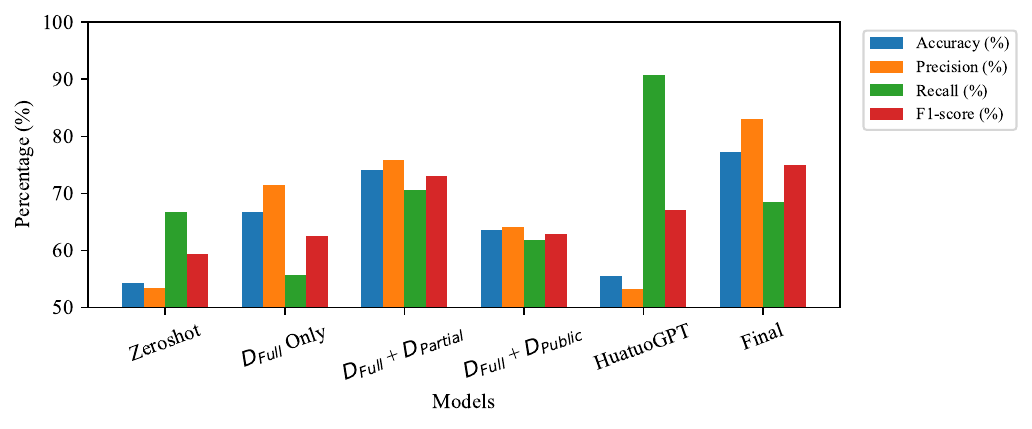}
        \caption{\textbf{Binary Classification Results on D\textsubscript{test}.} This bar chart shows the classification performance comparison of different model variants, including Accuracy, Precision, Recall, and F1-score. The models from left to right are: Zeroshot,  D\textsubscript{Full} Only,  D\textsubscript{Full} +  D\textsubscript{Partial},  D\textsubscript{Full} +  D\textsubscript{Public}, HuatuoGPT, and Final. The results demonstrate the effectiveness of multi-source data integration.}
        \label{Classification result}
    \end{figure}

\begin{figure}[htbp]
        \centering
        \includegraphics[width=1\linewidth]{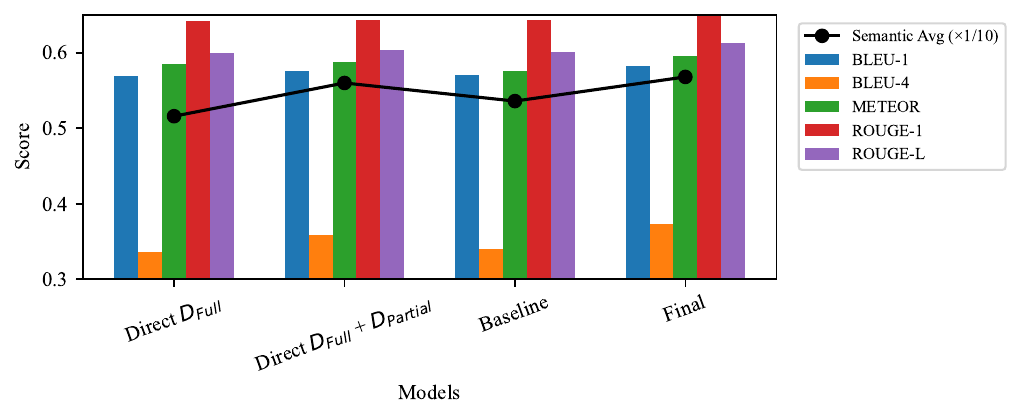}
        \caption{\textbf{Evaluation of Captioning Models on D\textsubscript{test}.} This figure shows the performance comparison of different model variants using Semantic Avg, BLEU-1, BLEU-4, METEOR, ROUGE-1, and ROUGE-L metrics. The models from left to right are: Direct D\textsubscript{Full}, Direct D\textsubscript{Full} + D\textsubscript{Partial}, Baseline, and Final. The final model achieves the best performance across most metrics.}
        \label{Caption result}
    \end{figure}

    Furthermore, the Stage 2 captioning performance (Figure~\ref{Caption result})
    highlights the success of our two-stage approach and similarity-guided
    pseudo-captioning. Initializing from the Stage 1 classification model and augmenting
    with pseudo-captions from D\textsubscript{Partial} led to the best lexical (e.g.,
    0.5953 METEOR) and semantic (6.36 DeepSeek-V3 Avg.) scores. This indicates that
    the foundational visual understanding from Stage 1 provides a strong basis for
    generating clinically relevant descriptions, and the pseudo-captions
    effectively transfer descriptive knowledge, enhancing caption quality in terms
    of completeness, professionalism, and accuracy without requiring extensive manual
    annotation. The poor performance of direct training and the HuatuoGPT-Vision
    baseline in captioning further emphasizes the value of domain-specific, staged
    fine-tuning.

    Crucially, the unified evaluation (Table~\ref{tab:unified evaluation}) revealsa
    synergistic relationship between classification and captioning. The Stage 2
    final model, after being trained for caption generation, exhibited an
    improved F1-score (78.13\%) on the classification task compared to the Stage
    1 model (75.06\%). This suggests that learning to generate detailed medical descriptions
    enhances the model's visual-semantic alignment and its ability to discern subtle
    diagnostic features, leading to a more robust and holistic understanding. Therefore,
    OralGPT demonstrates the potential to create a single, powerful vision-language
    model capable of performing both accurate automated diagnosis and generating
    coherent, clinically meaningful explanations, paving the way for advanced AI-assisted
    tools in oral healthcare that can support clinicians by not only identifying
    diseases but also by articulating the visual evidence.

\section{Conclusion}

In this paper, we introduced \textbf{OralGPT}, the first two-stage vision-language framework specifically designed for the diagnosis and description of oral mucosal diseases. We addressed the critical challenge of data scarcity in this specialized medical domain by constructing a comprehensive multi-source dataset and developing a novel similarity-guided pseudo-captioning method. This approach effectively leverages weakly-labeled data to generate high-quality supervision signals, enhancing both classification and captioning capabilities.

Our experimental results demonstrate the effectiveness of the OralGPT framework. The model achieved high accuracy in disease classification, outperforming baseline approaches by successfully integrating diverse data sources. Furthermore, it generated clinically relevant and fluent medical descriptions that captured key lesion characteristics, as validated by both lexical and advanced semantic metrics. A key finding is the synergistic relationship between the two stages: training the model to generate detailed captions (Stage 2) not only improved its descriptive abilities but also enhanced its diagnostic performance, indicating a more profound visual-semantic understanding.

The success of OralGPT lays a strong foundation for the development of advanced AI-assisted tools in oral healthcare. By providing both an accurate diagnosis and an interpretable, human-readable explanation, such models have the potential to support clinical decision-making, aid in dental education, and improve patient outcomes.

Future work will focus on expanding the dataset to include a wider range of diseases and data from multiple clinical centers to improve generalization. We also plan to conduct prospective clinical studies to evaluate the real-world performance and utility of OralGPT in a clinical setting. Exploring the integration of other data modalities, such as patient electronic health records, could further enhance the model's diagnostic and reasoning capabilities.

    \bibliographystyle{elsarticle-num}
    \bibliography{ref}

\begin{thebibliography}{10}
\expandafter\ifx\csname url\endcsname\relax
  \def\url#1{\texttt{#1}}\fi
\expandafter\ifx\csname urlprefix\endcsname\relax\def\urlprefix{URL }\fi
\expandafter\ifx\csname href\endcsname\relax
  \def\href#1#2{#2} \def\path#1{#1}\fi

\bibitem{intro1}
M.~Bankvall, E.~Dabelsteen, P.~Holmstrup, A.~C. Johannessen, M.~Jontell, E.~Neppelberg, J.~Rautava, Common oral mucosal lesions, Den norske tannlegeforenings Tidende 134~(2) (2024) 126--138.

\bibitem{intro2}
D.~A. Randall, N.~L.~W. Westmark, B.~W. Neville, Common oral lesions, American family physician 105~(4) (2022) 369--376.

\bibitem{neville2023oral}
B.~W. Neville, D.~D. Damm, C.~M. Allen, A.~C. Chi, Oral and maxillofacial pathology-E-Book, Elsevier Health Sciences, 2023.

\bibitem{warnakulasuriya2021oral}
S.~Warnakulasuriya, O.~Kujan, J.~M. Aguirre-Urizar, J.~V. Bagan, M.~{\'A}. Gonz{\'a}lez-Moles, A.~R. Kerr, G.~Lodi, F.~W. Mello, L.~Monteiro, G.~R. Ogden, et~al., Oral potentially malignant disorders: A consensus report from an international seminar on nomenclature and classification, convened by the who collaborating centre for oral cancer, Oral diseases 27~(8) (2021) 1862--1880.

\bibitem{lind1985malignant}
P.~O. Lind, H.~S. Koppang, E.~Aas, Malignant transformation in oral lichen planus, International journal of oral surgery 14~(6) (1985) 509--516.

\bibitem{bombeccari2011oral}
G.~P. Bombeccari, G.~Guzzi, M.~Tettamanti, A.~B. Giann{\`\i}, A.~Baj, F.~Pallotti, F.~Spadari, Oral lichen planus and malignant transformation: a longitudinal cohort study, Oral Surgery, Oral Medicine, Oral Pathology, Oral Radiology, and Endodontology 112~(3) (2011) 328--334.

\bibitem{intro3}
E.~A. Bilodeau, R.~V. Lalla, Recurrent oral ulceration: Etiology, classification, management, and diagnostic algorithm, Periodontology 2000 80~(1) (2019) 49--60.

\bibitem{zhang2022contrastive}
Y.~Zhang, H.~Jiang, Y.~Miura, C.~D. Manning, C.~P. Langlotz, Contrastive learning of medical visual representations from paired images and text, in: Machine learning for healthcare conference, PMLR, 2022, pp. 2--25.

\bibitem{liu2023visual}
H.~Liu, C.~Li, Q.~Wu, Y.~J. Lee, Visual instruction tuning, Advances in neural information processing systems 36 (2023) 34892--34916.

\bibitem{chen2024internvl}
Z.~Chen, J.~Wu, W.~Wang, W.~Su, G.~Chen, S.~Xing, M.~Zhong, Q.~Zhang, X.~Zhu, L.~Lu, et~al., Internvl: Scaling up vision foundation models and aligning for generic visual-linguistic tasks, in: Proceedings of the IEEE/CVF conference on computer vision and pattern recognition, 2024, pp. 24185--24198.

\bibitem{radford2021learning}
A.~Radford, J.~W. Kim, C.~Hallacy, A.~Ramesh, G.~Goh, S.~Agarwal, G.~Sastry, A.~Askell, P.~Mishkin, J.~Clark, et~al., Learning transferable visual models from natural language supervision, in: International conference on machine learning, PmLR, 2021, pp. 8748--8763.

\bibitem{han2020augmented}
S.~S. Han, I.~Park, S.~E. Chang, W.~Lim, M.~S. Kim, G.~H. Park, J.~B. Chae, C.~H. Huh, J.-I. Na, Augmented intelligence dermatology: deep neural networks empower medical professionals in diagnosing skin cancer and predicting treatment options for 134 skin disorders, Journal of Investigative Dermatology 140~(9) (2020) 1753--1761.

\bibitem{pelka2018radiology}
O.~Pelka, S.~Koitka, J.~R{\"u}ckert, F.~Nensa, C.~M. Friedrich, Radiology objects in context (roco): a multimodal image dataset, in: Intravascular Imaging and Computer Assisted Stenting and Large-Scale Annotation of Biomedical Data and Expert Label Synthesis: 7th Joint International Workshop, CVII-STENT 2018 and Third International Workshop, LABELS 2018, Held in Conjunction with MICCAI 2018, Granada, Spain, September 16, 2018, Proceedings 3, Springer, 2018, pp. 180--189.

\bibitem{gulshan2016development}
V.~Gulshan, L.~Peng, M.~Coram, M.~C. Stumpe, D.~Wu, A.~Narayanaswamy, S.~Venugopalan, K.~Widner, T.~Madams, J.~Cuadros, et~al., Development and validation of a deep learning algorithm for detection of diabetic retinopathy in retinal fundus photographs, jama 316~(22) (2016) 2402--2410.

\bibitem{de2018clinically}
J.~De~Fauw, J.~R. Ledsam, B.~Romera-Paredes, S.~Nikolov, N.~Tomasev, S.~Blackwell, H.~Askham, X.~Glorot, B.~O’Donoghue, D.~Visentin, et~al., Clinically applicable deep learning for diagnosis and referral in retinal disease, Nature medicine 24~(9) (2018) 1342--1350.

\bibitem{heo2021artificial}
M.-S. Heo, J.-E. Kim, J.-J. Hwang, S.-S. Han, J.-S. Kim, W.-J. Yi, I.-W. Park, Artificial intelligence in oral and maxillofacial radiology: what is currently possible?, Dentomaxillofacial Radiology 50~(3) (2021) 20200375.

\bibitem{hung2019application}
M.~Hung, M.~W. Voss, M.~N. Rosales, W.~Li, W.~Su, J.~Xu, J.~Bounsanga, B.~Ruiz-Negr{\'o}n, E.~Lauren, F.~W. Licari, Application of machine learning for diagnostic prediction of root caries, Gerodontology 36~(4) (2019) 395--404.

\bibitem{zhu2023artificial}
J.~Zhu, Z.~Chen, J.~Zhao, Y.~Yu, X.~Li, K.~Shi, F.~Zhang, F.~Yu, K.~Shi, Z.~Sun, et~al., Artificial intelligence in the diagnosis of dental diseases on panoramic radiographs: a preliminary study, BMC Oral Health 23~(1) (2023) 358.

\bibitem{khanagar2021developments}
S.~B. Khanagar, A.~Al-Ehaideb, P.~C. Maganur, S.~Vishwanathaiah, S.~Patil, H.~A. Baeshen, S.~C. Sarode, S.~Bhandi, Developments, application, and performance of artificial intelligence in dentistry--a systematic review, Journal of dental sciences 16~(1) (2021) 508--522.

\bibitem{uthoff2018point}
R.~D. Uthoff, B.~Song, S.~Sunny, S.~Patrick, A.~Suresh, T.~Kolur, G.~Keerthi, O.~Spires, A.~Anbarani, P.~Wilder-Smith, et~al., Point-of-care, smartphone-based, dual-modality, dual-view, oral cancer screening device with neural network classification for low-resource communities, PloS one 13~(12) (2018) e0207493.

\bibitem{rasteau2022artificial}
S.~Rasteau, D.~Ernenwein, C.~Savoldelli, P.~Bouletreau, Artificial intelligence for oral and maxillo-facial surgery: A narrative review, Journal of stomatology, oral and maxillofacial surgery 123~(3) (2022) 276--282.

\bibitem{lopez2022machine}
X.~A. L{\'o}pez-Cort{\'e}s, F.~Matamala, B.~Venegas, C.~Rivera, Machine-learning applications in oral cancer: A systematic review, Applied Sciences 12~(11) (2022) 5715.

\bibitem{aubreville2017automatic}
M.~Aubreville, C.~Knipfer, N.~Oetter, C.~Jaremenko, E.~Rodner, J.~Denzler, C.~Bohr, H.~Neumann, F.~Stelzle, A.~Maier, Automatic classification of cancerous tissue in laserendomicroscopy images of the oral cavity using deep learning, Scientific reports 7~(1) (2017) 11979.

\bibitem{cheng2016computer}
J.-Z. Cheng, D.~Ni, Y.-H. Chou, J.~Qin, C.-M. Tiu, Y.-C. Chang, C.-S. Huang, D.~Shen, C.-M. Chen, Computer-aided diagnosis with deep learning architecture: applications to breast lesions in us images and pulmonary nodules in ct scans, Scientific reports 6~(1) (2016) 24454.

\bibitem{alrashdan2016oral}
M.~S. Alrashdan, N.~Cirillo, M.~McCullough, Oral lichen planus: a literature review and update, Archives of dermatological research 308 (2016) 539--551.

\bibitem{keser2023deep}
G.~Keser, {\.I}.~{\c{S}}. Bayrakdar, F.~N. Pekiner, {\"O}.~{\c{C}}elik, K.~Orhan, A deep learning algorithm for classification of oral lichen planus lesions from photographic images: A retrospective study, Journal of stomatology, oral and maxillofacial surgery 124~(1) (2023) 101264.

\bibitem{zhou2024deep}
M.~Zhou, W.~Jie, F.~Tang, S.~Zhang, Q.~Mao, C.~Liu, Y.~Hao, Deep learning algorithms for classification and detection of recurrent aphthous ulcerations using oral clinical photographic images, Journal of Dental Sciences 19~(1) (2024) 254--260.

\bibitem{su2025deep}
A.-Y. Su, M.-L. Wu, Y.-H. Wu, Deep learning system for the differential diagnosis of oral mucosal lesions through clinical photographic imaging, Journal of Dental Sciences 20~(1) (2025) 54--60.

\bibitem{peng2024oral}
J.~Peng, Z.~Xu, H.~Dan, J.~Li, J.~Wang, X.~Luo, H.~Xu, X.~Zeng, Q.~Chen, Oral epithelial dysplasia detection and grading in oral leukoplakia using deep learning, BMC Oral Health 24~(1) (2024) 434.

\bibitem{adeoye2024deep}
J.~Adeoye, A.~Chaurasia, A.~Akinshipo, I.~Suleiman, L.-W. Zheng, A.~Lo, J.~Pu, S.~Bello, F.~Oginni, E.~Agho, et~al., A deep learning system to predict epithelial dysplasia in oral leukoplakia, Journal of Dental Research 103~(12) (2024) 1218--1226.

\bibitem{masood2024optimized}
H.~Masood, A.~Naseer, M.~Saeed, Optimized skin lesion segmentation: Analysing deeplabv3+ and assp against generative ai-based deep learning approach, Foundations of Science (2024) 1--25.

\bibitem{zhang2024research}
R.~Zhang, M.~Lu, J.~Zhang, X.~Chen, F.~Zhu, X.~Tian, Y.~Chen, Y.~Cao, Research and application of deep learning models with multi-scale feature fusion for lesion segmentation in oral mucosal diseases, Bioengineering 11~(11) (2024) 1107.

\bibitem{ramesh2025artificial}
E.~Ramesh, A.~Ganesan, K.~C. Lakshmi, P.~M. Natarajan, Artificial intelligence—based diagnosis of oral leukoplakia using deep convolutional neural networks xception and mobilenet-v2, Frontiers in Oral Health 6 (2025) 1414524.

\bibitem{alayrac2022flamingo}
J.-B. Alayrac, J.~Donahue, P.~Luc, A.~Miech, I.~Barr, Y.~Hasson, K.~Lenc, A.~Mensch, K.~Millican, M.~Reynolds, et~al., Flamingo: a visual language model for few-shot learning, Advances in neural information processing systems 35 (2022) 23716--23736.

\bibitem{boecking2022making}
B.~Boecking, N.~Usuyama, S.~Bannur, D.~C. Castro, A.~Schwaighofer, S.~Hyland, M.~Wetscherek, T.~Naumann, A.~Nori, J.~Alvarez-Valle, et~al., Making the most of text semantics to improve biomedical vision--language processing, in: European conference on computer vision, Springer, 2022, pp. 1--21.

\bibitem{li2023llava}
C.~Li, C.~Wong, S.~Zhang, N.~Usuyama, H.~Liu, J.~Yang, T.~Naumann, H.~Poon, J.~Gao, Llava-med: Training a large language-and-vision assistant for biomedicine in one day, Advances in Neural Information Processing Systems 36 (2023) 28541--28564.

\bibitem{lin2023pmc}
W.~Lin, Z.~Zhao, X.~Zhang, C.~Wu, Y.~Zhang, Y.~Wang, W.~Xie, Pmc-clip: Contrastive language-image pre-training using biomedical documents, in: International Conference on Medical Image Computing and Computer-Assisted Intervention, Springer, 2023, pp. 525--536.

\bibitem{chen2024huatuogpt}
J.~Chen, C.~Gui, R.~Ouyang, A.~Gao, S.~Chen, G.~H. Chen, X.~Wang, R.~Zhang, Z.~Cai, K.~Ji, et~al., Huatuogpt-vision, towards injecting medical visual knowledge into multimodal llms at scale, arXiv preprint arXiv:2406.19280 (2024).

\bibitem{zhou2023skingpt}
J.~Zhou, X.~He, L.~Sun, J.~Xu, X.~Chen, Y.~Chu, L.~Zhou, X.~Liao, B.~Zhang, X.~Gao, Skingpt-4: an interactive dermatology diagnostic system with visual large language model, arXiv preprint arXiv:2304.10691 (2023).

\bibitem{yan2025multimodal}
S.~Yan, Z.~Yu, C.~Primiero, C.~Vico-Alonso, Z.~Wang, L.~Yang, P.~Tschandl, M.~Hu, L.~Ju, G.~Tan, et~al., A multimodal vision foundation model for clinical dermatology, Nature Medicine (2025) 1--12.

\bibitem{wang2023chatcad}
S.~Wang, Z.~Zhao, X.~Ouyang, Q.~Wang, D.~Shen, Chatcad: Interactive computer-aided diagnosis on medical image using large language models, arXiv preprint arXiv:2302.07257 (2023).

\bibitem{zhu2023minigpt}
D.~Zhu, J.~Chen, X.~Shen, X.~Li, M.~Elhoseiny, Minigpt-4: Enhancing vision-language understanding with advanced large language models, arXiv preprint arXiv:2304.10592 (2023).

\bibitem{moor2023med}
M.~Moor, Q.~Huang, S.~Wu, M.~Yasunaga, Y.~Dalmia, J.~Leskovec, C.~Zakka, E.~P. Reis, P.~Rajpurkar, Med-flamingo: a multimodal medical few-shot learner, in: Machine Learning for Health (ML4H), PMLR, 2023, pp. 353--367.

\bibitem{sun2023evaluating}
Z.~Sun, H.~Ong, P.~Kennedy, L.~Tang, S.~Chen, J.~Elias, E.~Lucas, G.~Shih, Y.~Peng, Evaluating gpt-4 on impressions generation in radiology reports, Radiology 307~(5) (2023) e231259.

\bibitem{liu2025radiology}
Z.~Liu, Y.~Li, P.~Shu, A.~Zhong, H.~Jiang, Y.~Pan, L.~Yang, C.~Ju, Z.~Wu, C.~Ma, et~al., Radiology-gpt: a large language model for radiology, Meta-Radiology (2025) 100153.

\bibitem{jiang2022two}
L.~Jiang, D.~Chen, Z.~Cao, F.~Wu, H.~Zhu, F.~Zhu, A two-stage deep learning architecture for radiographic staging of periodontal bone loss, BMC Oral Health 22~(1) (2022) 106.

\bibitem{krois2019deep}
J.~Krois, T.~Ekert, L.~Meinhold, T.~Golla, B.~Kharbot, A.~Wittemeier, C.~D{\"o}rfer, F.~Schwendicke, Deep learning for the radiographic detection of periodontal bone loss, Scientific reports 9~(1) (2019) 8495.

\bibitem{chang2020deep}
H.-J. Chang, S.-J. Lee, T.-H. Yong, N.-Y. Shin, B.-G. Jang, J.-E. Kim, K.-H. Huh, S.-S. Lee, M.-S. Heo, S.-C. Choi, et~al., Deep learning hybrid method to automatically diagnose periodontal bone loss and stage periodontitis, Scientific reports 10~(1) (2020) 7531.

\bibitem{he2020pathvqa}
X.~He, Y.~Zhang, L.~Mou, E.~Xing, P.~Xie, Pathvqa: 30000+ questions for medical visual question answering, arXiv preprint arXiv:2003.10286 (2020).

\bibitem{van2023clinical}
D.~Van~Veen, C.~Van~Uden, L.~Blankemeier, J.-B. Delbrouck, A.~Aali, C.~Bluethgen, A.~Pareek, M.~Polacin, E.~P. Reis, A.~Seehofnerova, et~al., Clinical text summarization: adapting large language models can outperform human experts, Research square (2023) rs--3.

\bibitem{xie2020self}
Q.~Xie, M.-T. Luong, E.~Hovy, Q.~V. Le, Self-training with noisy student improves imagenet classification, in: Proceedings of the IEEE/CVF conference on computer vision and pattern recognition, 2020, pp. 10687--10698.

\bibitem{sohn2020fixmatch}
K.~Sohn, D.~Berthelot, N.~Carlini, Z.~Zhang, H.~Zhang, C.~A. Raffel, E.~D. Cubuk, A.~Kurakin, C.-L. Li, Fixmatch: Simplifying semi-supervised learning with consistency and confidence, Advances in neural information processing systems 33 (2020) 596--608.

\bibitem{vanrijsbergen1979}
C.~J. van Rijsbergen, Information Retrieval, Butterworth-Heinemann, 1979.

\bibitem{papineni2002bleu}
K.~Papineni, S.~Roukos, T.~Ward, W.-J. Zhu, Bleu: a method for automatic evaluation of machine translation, in: Proceedings of the 40th annual meeting of the Association for Computational Linguistics, 2002, pp. 311--318.

\bibitem{banerjee2005meteor}
S.~Banerjee, A.~Lavie, Meteor: An automatic metric for mt evaluation with improved correlation with human judgments, in: Proceedings of the acl workshop on intrinsic and extrinsic evaluation measures for machine translation and/or summarization, 2005, pp. 65--72.

\bibitem{lin2004rouge}
C.-Y. Lin, Rouge: A package for automatic evaluation of summaries, in: Text summarization branches out, 2004, pp. 74--81.

\end{thebibliography}
\end{document}